%% file: mcsesans_X.tex
\documentclass{article}

\usepackage{amssymb}
\usepackage{graphicx}

\newcommand{\txt}[1]{{\mbox{\tiny #1}}}
\newcommand{\bracket}[1]{\left(#1\right)}

\newcommand{\angular}[1]{\left[#1\right]}

\newcommand{\eqn}[1]{(\ref{#1})}
\renewcommand{\d}[1]{(\mathrm{d})}
\newcommand{\etal}[1]{\emph{et al.\ }}

\title{Monte Carlo Calculation of the Single-Particle Spin-Echo
Small-Angle Neutron Scattering Correlation Function}

\author{H\aa kon Kaya\\
Department of Chemical Engineering, University of Amsterdam,\\ 
Nieuwe Achtergracht 166, 1018 WV Amsterdam, The Netherlands.\\
hkaya@ulb.ac.be}

\begin{document}
\maketitle

\begin{abstract}
A Monte Carlo algorithm for calculating the single-particle spin-echo
small-angle neutron scattering (SESANS) correlation function is
presented. It is argued that the algorithm provides a general and
efficient way of calculating SESANS data for any given shape and
structure. 
\end{abstract}



\clearpage

\input{mcsesans_intro}

\input{mcsesans_theo}

\input{mcsesans_mc}

\input{mcsesans_res}

\begin{appendix}
\input{cumuldist}
\end{appendix}

\bibliography{SESANS}

\input{mcsesans_figtabs}

\end{document}

%% file: mcsesans_intro.tex
\section{Introduction}

Spin-echo small-angle neutron scattering (SESANS) has recently emerged
as a new way of applying neutron scattering to the investigation of
the structure of matter \cite{Wim2000,MTR2000}. The method is
particularly useful for large structures in the size range from 10 nm
up to several microns. This is the same size range covered by
techniques like light scattering and ultra-small angle neutron
scattering (USANS). The use of neutron spin echo in measuring
elastic scattering, however, renders beam collimation unnecessary,
thus avoiding the low fluxes from which USANS suffers. In comparison
to light scattering, the use of neutron allows for study of opaque or
highly concentrated samples. 
The SESANS method is presently on an active developing stage, 
and the theoretical concepts and methods from which conventional
small-angle neutron scattering (SANS) benefits have only recently
started to be derived \cite{Tim2003} and be applied to the analysis 
of experimental data \cite{Tim2003b}. 
As in the case of SANS, the data analysis can be 
performed with model-dependent or model-independent methods. 
In the latter case, the experimentally obtained scattering functions
are inverted to obtain a curve representing the pair distance
distribution function $p(r)$. The most well-known realisation for
this procedure is the Indirect Fourier Transform by 
Glatter \cite{Glatter77a}. 
Similar analysis can be carried out by the
maximum entropy method \cite{Tsao} and the regularization method of 
Svergun \cite{Svergun}.  
Model-independent analysis is particularly simple in
the case of SESANS, as the relation between the SESANS correlation
function $G(z)$ and the small-angle scattering correlation function
$\gamma(r)$ is given by an Abel integral equation \cite{Tim2003b}, for
which there exist standard numerical methods for solution.

In model-dependent analysis, mathematical functions that model the
scattering intensity from a system of particles with presumed shape,
structure, and ordering are fitted to the experimental data. 
From the fitted parameters one obtains information such as the size
and shape of the particles, their inner structure and size 
distribution, and the inter-particle interactions that create ordered
structures. 
In SANS terminology, the last piece of information is contained in the
\textit{structure factor}, whereas information pertaining to
single-particle scattering is contained in the \textit{form factor}.
Analytical or semi-analytical functions for the scattering
form factors or scattering amplitudes are known for several geometries
\cite{Pedersen97}. These functions are easy to extend to 
include core-shell structures and polydisperse assemblies.

As will be elaborated in the following section, 
$G(z)$ is related to the SANS scattering cross section
$\d\Sigma/\d\Omega$ by a two-dimensional cosine transform
\cite{Uca2003}. Knowing the full detectable $\d\Sigma/\d\Omega$ 
as an analytical function or as tabulated values of
$\d\Sigma/\d\Omega$ vs.\ $Q$, calculation of 
$G(z)$ is straightforward by numerical integration.
%
In principle, it would be desirable and more efficient with analytical
expressions for $G(z)$ for different geometries, analogous to form
factors for SANS, so that SESANS data can be analyzed with a similar 
tool-box of model functions. 
Analytical expressions for $G(z)$ for scattering from single
homogenous and hollow spheres have already been derived using the
concept of the mass correlation function $\gamma(r)$
\cite{Tim2003}. If $\gamma(r)$ of a given structure is known,
calculation of $G(z)$ is simple. 
For non-spherical geometries, however, expressions for $\gamma(r)$ can
take complicated forms \cite{Hemisphere,Cone,Gille}. 
Moreover, it is not an easy task to extend $\gamma(r)$ from homogenous
to multi-domain structures. 

In this paper we investigate an alternative method to calculate
$G(z)$. It is for all practical purposes of general validity and
straightforward to implement for any shape and structure. 
The idea is to perform a Monte Carlo calculation of the pair
distance distribution function $p(r)$, from which the SESANS
correlation function $G(z)$ and also the SANS
scattering cross section $\d\Sigma/\d\Omega$ can be 
calculated by a single numerical integration.  
Monte Carlo methods have been used by several 
authors in calculation of SANS spectra
\cite{Hubbard88,Hansen90,McAlister98,Flavio2003}. 
The algorithm for calculating SESANS curves is outlined in Section 3. 
In Section 4 we present results of the calculations for different
shapes and structures. 

%% file: mcsesans_theo.tex
\section{SESANS Theory}

The measured quantity in a SESANS experiment is the loss of
polarization a neutron beam suffers by being scattered by the
sample. By passing through magnetic fields before and after
interacting with the sample, the neutrons are subjected to 
Larmour spin precession. In the case of no interaction with the
sample, the precessions before and after the sample area cancel each
other, yielding a spin echo preserving the polarization state of the
beam \cite{MTR2000}. 
The presence of a scattering sample produces precession lengths
differences that are functions of the scattering angle.
The ensuing depolarization is a function of the SESANS 
correlation function $G(z)$ \cite{Uca2003}:
\begin{equation}\label{eq:P}
\frac{P(z)}{P_0} = \exp\angular{G(z) - G(0)} 
  =  \exp\angular{G(0)\bracket{G_0(z) - 1} },
\end{equation}
where $G_0(z)$ is the normalized correlation function.
The relation between the SANS macroscopic scattering cross section  
$\d\Sigma/\d\Omega$ and the SESANS correlation function has
already been derived \cite{Wim2000,MTR2000}:  
\begin{equation}\label{eq:Gz_SANS}
G(z) = \frac{\lambda^2 t}{4\pi^2} 
	\int_{-\infty}^\infty \d Q_y 
	\int_{-\infty}^\infty \d Q_z
	\frac{\d\Sigma(\mathbf{Q})}{\d\Omega}\,
	\cos\bracket{Q_z z},
\end{equation}
where $\lambda$ and $t$ denote the wavelength of the neutron beam and
the thickness of the sample, respectively. 
$Q_y$ and $Q_z$ are the cartesian components of the scattering vector
$\mathbf{Q}$, the incident beam lying along the $x$ axis. 
The integrations in \eqn{eq:Gz_SANS} are in practice defined by the
area in the $yz$-plane covered by the detector. The spin-echo length
$z$ is a function of the neutron wavelength, the sample position, and
the configuration of the magnetic fields \cite{Wim2000}.

We now consider a system of non-interacting particles isotropically
embedded in a homogenous matrix or dispersed a solvent. The SANS
scattering cross section can be written in terms of an intra-particle
form factor $P(Q)$ and an inter-particle structure factor $S(Q)$:
\begin{equation}
\frac{\d\Sigma}{\d\Omega}(Q) = n_p V^2 P(Q) S^\prime(Q),
\end{equation}
where $n_p$ is the number density of scattering particles and $V$ 
is the volume of a particle.
Most of the analytical structure factors have been calculated for
systems of monodisperse spheres. The effective structure factor
$S^\prime(Q)$ includes approximate corrections to $S(Q)$ due to
particle polydispersity or anisotropy \cite{kc83,Pedersen94,bioscal}.
We will in the following consider dilute system, for which we may
ignore inter-particle scattering and set $S^\prime(Q) = 1$,
corresponding to an ideal gas. 
The form factor $P(Q)$ is related to the average size and
shape of the individual particles and to their inner structure. 
Focusing on a single particle, the form factor can be written in terms
of the density correlation function $\gamma(r)$:
\begin{equation}
P(Q) = \int_0^D 
\gamma(r) \frac{\sin Qr}{Qr}\,  4\pi r^2\d r,
\end{equation}
where $\gamma(r)$ is defined by \cite{Guinier,Glatter}:
\begin{equation}\label{eq:pr}
\gamma(r) = \frac{1}{V}\left< \int_V \d\mathbf{r}^\prime
\Delta\rho\bracket{\mathbf{r}^\prime} \,
\Delta\rho\bracket{\mathbf{r}^\prime+\mathbf{r}} \right>,
\end{equation}
where the braces $\left<\right>$ denote averaging over all
orientations of the position vector
$\mathbf{r}$. $\Delta\rho(\mathbf{r})$ 
is the scattering length density at a position $\mathbf{r}$ inside
the particle, minus the constant scattering length of the surrounding
medium (in most cases a solvent). $D$ is the largest chord length of
the particle, so that $\gamma(r) = 0$ for $r > D$.   
For a homogenous particle $\gamma(r)$ is proportional to the overlap
volume between the particle and its identical ``ghost'' that has been
shifted by a distance $r$. For an inhomogenous particle, the volume
of the overlapping region must be weighted with the product of the
scattering length densities of the respective regions
\cite{Henderson96}. An important identity is \cite{Glatter}:
\begin{equation}\label{eq:normgamma}
\bracket{\Delta\rho}^2 V = \int_0^D \gamma(r)\,4\pi r^2\d r,
\end{equation}
where $\Delta\rho$ is the difference between the average scattering
length density of the particle and that of the surrounding medium (in
most cases a solvent). The normalized density autocorrelation function
$\gamma_0(r)$ is defined through $\gamma(r) =
\bracket{\Delta\rho}^2\gamma_0(r)$ and has the property $\gamma_0(0) =
1$. 

%

In this paper we focus on the single-particle contribution to the
SESANS spectrum. A direct real-space interpretation of $G(z)$ was
presented by 
Krouglov \etal \cite{Tim2003}.
For a system of non-interacting particles, the SESANS
correlation function can be written $G(z) = \lambda^2t n_p 
G_p(z)$, where the single-particle SESANS correlation function bears
the following relation to the structure function $\gamma(r)$: 
\begin{eqnarray}
G_p(z) & = & V \int_{-\infty}^\infty \gamma\bracket{\sqrt{x^2 + z^2}}
	\d x  
\nonumber\\
 & = & 2 \bracket{\Delta\rho}^2 V
	\int_0^{\sqrt{D^2-z^2}} 
	\gamma_0\bracket{\sqrt{x^2 + z^2}} \d x. \label{eq:GXZ}
\end{eqnarray}
Note that both $G(z)$ and $G_p(z)$ are dimensionless. 
Knowing $\gamma(r)$, the SANS form factor and the single-particle
SESANS correlation function can be calculated. 
An important quantity in SESANS is the total scattering probability,
given as $G(0)$ \cite{Tim2003,Uca2003}. It relates to the observed 
depolarization and thus gives an indication on the magnitude and 
detectability of the SESANS signal. From the above we have:
\begin{eqnarray}
G(0) & = & 2\lambda^2t n_p \,
V \int_0^D \gamma\bracket{r } \d r \nonumber\\
& = & 2 \lambda^2 t\, \phi \bracket{\Delta\rho}^2
\int_0^D \gamma_0\bracket{r } \d r \\
& = & \lambda^2 t\, \phi \bracket{\Delta\rho}^2 \overline{l},
\label{eq:lbar}
\end{eqnarray}
where $\phi = n_p V$ is the volume fraction of the particles
and $\overline{l}$ 
is the mean length of all chords contained in the particle
\cite{Guinier}. 
Finally in this section, we remark that in the case of an ensemble of
polydisperse, non-interacting particles, the equations above
take the form
\begin{eqnarray}
G(z) & = & 2 \lambda^2t n_p 
\left< \bracket{\Delta\rho}^2 V
\int_0^{\sqrt{D^2-z^2}}\gamma_0\bracket{\sqrt{x^2+z^2}} \,\d x
\right> \label{eq:polyGz} \\
G(0) & = & 2 \lambda^2t \phi\,
\frac{1}{\left<V\right>} 
\left< \bracket{\Delta\rho}^2 V
\int_0^D \gamma_0\bracket{x}\, \d x \right> ,
\label{eq:polyG0} 
\end{eqnarray}
where $\left<\right>$ now stands for the averaging over the particle
sizes. The number density is given by $n_p = \phi/\left<V\right>$. 
If $\bracket{\Delta\rho}^2$ is the same for all particles, Equation
\eqn{eq:polyG0} can be written $G(0) = \lambda^2t \phi
\bracket{\Delta\rho}^2 \overline{l}_w$, where $\overline{l}_w$ is the
weight-averaged mean chord length.

%% file: mcsesans_mc.tex
\section{Monte Carlo calculation of $G(z)$}
\label{sec:montecarlo}

\subsection{Calculating the pair distance distribution function}

The pair correlation function $\gamma(r)$ is related to the pair
distance distribution function (pddf) $p(r)$ by 
\begin{equation}\label{eq:pr2g}
p(r) = r^2 \gamma(r).
\end{equation}
$p(r)$ is the probability of two random points within the particle
being separated by a distance $r$. By random sampling of distances
between points within the particle, and keeping statistics of the
sampled distances, $p(r)$ can be found for any particle shape. 
The interval $0 \le r \le D$ is partitioned into $M+1$
histogram bins, indexed from 0 to $M$, $D$ being the maximum distance
between two points belonging to the particle.
$M = 200$ was used for the calculations presented in this paper. 

Geometrical points are sampled uniformly from a volume that
circumscribes the volume of the given particle as closely as
possible. When $N_m=1000$ points have been selected, 
the points that fall outside the shape function of the particle
are discarded, leaving $N_r$ points. The points are sampled in batches
of $N_m$ in order not to exhaust the computer memory. 
The shape of the sampling volume is essential, as it ensures that most
of the sampled points will belong to the particle. 
This makes the calculations far more efficient than sampling from a
circumscribing rectangular box, discarding the points that fall
outside the shape function of the particle \cite{Hubbard88,McAlister98}.
The sampling volumes are rectangular, spherical, or cylindrical,
depending on the shape of the particle. 
Uniform sampling from a given distribution or volume
by an inverse method is a well-known technique in Monte Carlo  
calculations \cite{NumericalRecipes}. For completeness,
we give a brief outline of the procedure in Appendix
\ref{appendix:uniform_sampling}
For the spherical and
cylindrical sampling volumes, it is straightforward to limit the
samling to given segments or sectors. Regardless of the shape of the
simulation box, it is necessary that it encloses the entire particle
ensure that different parts of the particle contribute to $p(r)$ in
proportion to their volume. This can also be accomplished by
allocating a fraction $V_i/V$ of the randomly selected points to each
domain $i$. This alternative is particularly suited for
multidomain structures with inhomogenous density distributions.

The algorithms proceeds by calculating the $N_r(N_r-1)/2$ distances
defined by the accepted points. The distance 
$d = \sqrt{\left|\mathbf{r}_a - \mathbf{r}_b\right|^2}$ 
between the points $\mathbf{r}_a$
and $\mathbf{r}_b$ is counted into the histogram,
wheighted by the product of the scattering length densities $\rho_a$
and $\rho_b$ of the two points. 
The update of the histogram is carried out by the
following algorithm:     
\begin{equation}\label{eq:count}
\tilde{p}(i) \longleftarrow 
\tilde{p}\bracket{i = \mathrm{int} 
\bracket{\frac{Md}{D} + 0.5}} + \rho_a \rho_b,
\end{equation}
where the int() function represents truncation of decimals, leaving
the integer index $i$ of the histogram bin. The index $i$ is related
to the intraparticle distance by $r = i D/M$. 
Adding 0.5 to the argument of the int() eliminates the problem
overcounting smaller distances because of the decimal truncation. 

When the $N_r(N_r-1)/2$ distances have been counted, the program
checks the accuracy of the calculated $p(i)$ after a criterion to be
described below. If the accuracy is not accepted, an new batch of
$N_r$ points is selected, and the resulting new $N_r(N_r-1)$ distances
are counted into the histogram following equation \eqn{eq:count}. 
The total number of sampled points belonging to the particle, ie
accepted points, is denoted $N_p$; and the total number of sampled
points is denoted $N_M$.

The normalization of $p(i)$ is done so that the form factor
will satisfy $P(Q=0) = 1$. This is accomplished by calculating  
$p(i) = \tilde{p}(i) / \bracket{C D/M}$, where
\begin{equation}\label{eq:normal}
C = \sum_{a=1}^{N_p-1} \sum_{b=a+1}^{N_p} \rho_a \rho_b 
= \bracket{\Delta\rho}^2 N_p(N_p-1)/2 
\end{equation}
is the sum of the distance weights. The last equality in
\eqn{eq:normal} is valid in the limit of large $N_p$. For a homogenous
particle, $C$ will be proportional to the number of sampled
intra-particle distances. Knowing $C$, $\bracket{\Delta\rho}^2$ can
thus be calculated. In most cases the volume $V$ of the particle will
be known beforehand. If not, it can be found from the MC calculations
by the relation $V = N_p V_\txt{box}/N_M$, where $V_\txt{box}$ is the
volume of the simulation box. 

Numerical tests showed that a reliable test for the accuracy of $p(i)$
was to compare $p(i)_N$, the pair distance distribution 
function calculated from $N$ sampled points, with $p(i)_{N-1000}$.  
This was done for every time $p(i)$ had been calculated with 1000 
new points and upgraded with $1000\times(1000-1)/2$ distances. 
The calculations were halted when 
\begin{equation}
\sqrt{ \frac{\sum_i \angular{ p(i)_N - p(i)_{N-1000} }^2 }
{\sum_i \angular{ p(i)_{N-1000} }^2 } }  < 0.25\%.
\end{equation}
For homogenous particles, the required number of points lay around
$N_P=5000$. For inhomogenous particles, the number could be
significantly higher; for core-shell particles with equal volumes and
opposite signs of the scattering length densities, values up to
$N_P=40000$ were typical. Nevertheless, calculation of a full data set
$G_0(z)$ took only a few seconds on a notebook equipped with a
750 MHz Pentium-III processor.   

\subsection{Finding $\gamma(r)$}
\label{sec:findgamma} 

To find $\gamma(r)$ from the calculated $p(r)$, Eq.~\eqn{eq:pr2g}
faces us with the problem of dividing by zero or small values of
$r^2$. We overcome this problem by exploiting the small-$r$ expansion
of $\gamma(r)$:
\begin{equation}\label{eq:expgamma}
\gamma(r) = a + b r + c r^2 + \ldots
\end{equation}
The expansion parameters are known as \textit{differential parameters}
and are related to structural features of the particle
\cite{Glatter}. The second order parameter $c$ will in most cases be
zero, and the last term could thus be neglected or replaced with a
third-order term $d r^3$. For the sake of the generality of the
algorithm, however, we apply the expansion as given in equation
\eqn{eq:expgamma}. 
One should bear in mind, though, that there are particle shapes for 
which an expansion like \eqn{eq:expgamma} can not be carried out 
\cite{Cone}. However, it is valid for most realistic geometries.
We find the differental parameters $a, b, c$ by polynomial
least-squares fitting \cite{FairesBurden} of $ar^2 + br^3 + cr^4$ 
to the Monte Carlo calculated $p(r)$ at small $r$.
For the sum of squared residuals we have: 
\begin{equation}\label{eq:Em}
E_m(a, b, c) = \sum_{i=0}^m \angular{p_i 
- \bracket{ar_i^2 + br_i^3 + cr_i^4}}^2,
\end{equation}
where $r_i = i D/M$. $E_m$ is minimized with respect to
$a$, $b$, and $c$; and the resulting linear system is solved 
for $a, b, c$ with standard routines \cite{NumericalRecipes}. 
%
The summation in \eqn{eq:Em} runs from $i=0$ to $i=m$, where the index
number $m < M$ is decremented until the sum of squared residuals
$E_m(a, b)$ has a sufficiently low value. When this is the case,
$\gamma(r)$ up to index $i=m$ is given by $a + br_i + c r_i^2$. For
the remaining values of $r$, $\gamma(r)$ is calculated directly by
$p(r)/r^2$. An example is demonstrated in Figure \ref{fig:Pr_sphere}.
In the calculations in this paper, the initial value of $m$ 
is set at half the distance between $r=0$ and the first
peak of $p(r)$. Finally, to facilitate a consistent computation of
$G(z)$ and $G(0)$, $\gamma_0(r) = \gamma(r)/a$ is calculated by
normalization. 

\subsection{Calculation of $G(0)$ and $G_0(z)$} 
\label{sec:findGz}

The single-particle SESANS correlation function $G_p(z)$ is calculated
by numerical evaluation of the integral \eqn{eq:GXZ}, neglecting
the prefactor $2\bracket{\Delta\rho}^2V$.
Having found $\gamma_0(r)$, we
need to evaluate $\gamma_0(\sqrt{x^2+z^2})$ for arbitrary values of
$x$ and $z$. This is accomplished by natural cubic spline
interpolation \cite{NumericalRecipes}. The integral \eqn{eq:GXZ} is
evaluated using a 16-point Gauss-Legendre quadrature 
\cite{NumericalRecipes,FairesBurden}. 
The normalization of $G_p(z)$ is carried out by dividing by $G_p(0)$,
which is calculated at the beginning by evaluating \eqn{eq:GXZ} for
$z=0$. Subsequent calculations of $G_p(z)$ are normalized by dividing
by $G_p(0)$, thus yielding $G_0(z)$. The total scattering probability
$G(0)$ is given by $2\bracket{\Delta\rho}^2 n_p \lambda^2 t G_p(0)$,
where the calculation of $\bracket{\Delta\rho}^2$, if unknown a
priori, can be done by means of Equation \eqn{eq:normal}.

%% file: mcsesans_res.tex
\section{Results and discussion}

In Figure \ref{fig:sphere} we have plotted the MC-calculated $G_0(z)$
curve for a homogenous sphere with radius $R=50$ together with the
analytical expression for $G_0(z)$ \cite{Tim2003}. The latter reads: 
\begin{eqnarray}
G_0(\zeta) & = & \bracket{1 - \bracket{\frac{\zeta}{2}}^2}^{1/2} 
\bracket{1 + \frac{\zeta^2}{8}} \nonumber\\
& & + \frac{\zeta^2}{2} \bracket{1 - \bracket{\frac{\zeta}{4}}^2}
\ln \bracket{ \frac{\zeta}{2 + \sqrt{4 -\zeta^2}} },
\label{eq:Gz_sphere}
\end{eqnarray}
where $\zeta = z/R$. The agreement between the calculated data and the
analytical function is excellent. The other curves represent 
MC-calculated $G_0(z)$ for spheres with $R=50$, but with radial
density profiles following a hyperbolic form $\rho(r) = r^{-\alpha}$. 
The density profile has a pronounced effect on the appearance of the
$G_0(z)$ curves and also on the total scattering probability $G(0)$. 
For the full sphere, one has $\overline{l} = 3R/2 = 75$, whereas 
for $\rho(r) = r^{-2}$, the calculations gave $\overline{l} = 5.4$.  

Figure \ref{fig:hsphere} shows the MC-calculated $G_0(z)$ for hollow
spheres with outer radius $R=50$ and varying inner radii. Comparing
$G_0(z)$ of the hollow spheres with that of the full sphere, the
interesting feature is the appearance of a small shoulder at a
$z$-value corresponding to the inner diameter of the hollow spheres. 

$G_0(z)$ for core-shell spheres with inner radius $R_1=30$ and outer
radius $R_2=50$ are plotted in Figure \ref{fig:coreshell_sphere} for
different combinations of the scattering length densities $\rho_1$ and
$\rho_2$. Some values of $\rho_1, \rho_2$ give rise to strong
oscillations in $G_0(z)$, which were also found in the multishell
calculations by 
Rekveld \etal \cite{MTR2000}.
The quantitative interpretation
of these oscillations was given by 
Uca \cite{Uca2003}.
Minima arise from correlations between particle regions with opposite
signs of their scattering length densities. Correspondingly, maxima
are related to identical signs in different regions, or to correlations
within the same region. The positions of these extrema give information
on the typical distances between or within these regions.
Thus, the position of minimum at $z=40$ in Figure
\ref{fig:coreshell_sphere} is related to the typical shell-core
distance, which is taken as the core radius plus half the shell
thickness, which in the case shown in the Figure is exactly $R_1 +
\bracket{R_2-R_1}/2 = 40$. The maximum at $z\approx70$ is due to the
shell-shell correlations, of which the typical distance, taken as the
core diameter plus twice the half shell thickness, is 80. The
core-core correlations are expected to give a maximum around
$z=R_1=30$, but this is hidden by the inital part of $G_0(z)$. 

Alternatively, oscillations in $G_0(z)$ can be interpreted in terms of
the differential parameters introduced in Section \ref{sec:findgamma}.
Inserting the expansion \eqn{eq:expgamma} into \eqn{eq:GXZ} shows that
each term in \eqn{eq:expgamma}, except the zero-order term $a$,
produces a contribution to $G_0(z)$ containing a maximum. The position
such a maximum is shifted to higher $z$ for higher order terms. Thus
the minimum of $G_0(z)$ at $z=40$ in Figure \ref{fig:coreshell_sphere}
can be attributed to a large negative differential parameter of high
order in the expansion \eqn{eq:expgamma}. Accordingly, the maximum at
$z=70$ comes from a positive differential parameter of even higher
order. Relating such high-order differential parameters directly to
structural features of the particle in a unique way is, however, a
challenging task in small-angle scattering theory
\cite{Glatter,Ciccariello88,Ciccariello91}.  

MC-calculated $G_0(z)$ for homogenous and hollow cylinders are plotted
in Figure \ref{fig:cylinder}. It should be noted that for cylinders
with a high aspect ratio $L/2R$, the maximum of $p(r)$ is shifted
towards small $r$. As a consequence, the least-squares method for
determining $\gamma(r)$ described in Section \ref{sec:findgamma} must
use a small number of data points. However, this is not found to pose
a serious problem.  
The $G_0(z)$ curve for the homogenous cylinder shows the same 
characteristic features as in the model calculations by
Uca \etal \cite{Uca2003}. 
For the hollow cylinder, however, there are
remarkable differences. The SESANS functions are concave at small $z$,
and there is a clear shoulder at a position corresponding to the inner
diameter. For the thinnest of the cylindrical shells, the shoulder
appears to give a discontinous first derivative at $z=100$. In the
corresponding $p(r)$ curve, the maximum also appeared
discontinous. Because of this, it was necessary to increase the
accuracy of the numerical quadrature described in Section
\ref{sec:findGz} to avoid numerical artefacts. 
At the highest $z$-values, the curves practically coincide, all
showing the loss of correlations characteristic for anisotropic
structures. 

SESANS functions for core-shell cylinders analogous to the core shell
spheres in Figure \ref{fig:coreshell_sphere} are plotted in Figure 
\ref{fig:coreshell_cylinder}. The inner and outer radii are $R_1 = 30$
and $R_2 = 50$, and the cylinder length is $L=250$. 
The positions of minima and maxima can be interpreted in the same way
as for the spheres. For the lowest curve, the correlations are almost
completely lost already at $z=100$. This can be related to the fact
that for this particular combination of core and shell volumes and
scattering length densities, the overall scattering length densities
of the core and shell are nearly equal, but of different signs. Thus
for $z$ greater than 100, corresponding to correlations only along the
cylinder axis, the core and shell contributions cancel each other out.

In Figure \ref{fig:ellipsoids} the SESANS functions for various
triaxial ellipsoids are shown. The case for ellipsoids of revolution
have been discussed by 
Uca \etal \cite{Uca2003}.
The SESANS functions of the ellipsoids with small eccentricities have
been calculated by sampling points from a circumscribing sphere. For
higher eccentricities, a circumscribing cylinder was used.
Sampling random points from a confocal ellipsoidal coordinate
system \cite{wolframEI} would in principle eliminate the problem of
sampling and discarding points that fall outside the
ellipsoid. However, applying the inversion method of Appendix
\ref{appendix:uniform_sampling} to such special coordinate systems
would in most cases require the numerical solution of nonlinear
equations, thus loosing the advantage of sampling efficiency.

We close this section with a discussion of the efficiency of the
Monte Carlo method, with the prospective application of analyzing real
SESANS data. 
The calculations in this work were performed using the random number
generation \verb ran3  given by 
Press \etal \cite{NumericalRecipes}.
This is a very reliable routine, but for the purpose of rapid
calculations, simpler generators could be applied.
Although the Monte Carlo method is general, the examples shown are for
relatively simple structures. 
Additional refinements are possible, but to the cost of increased
computing time.  
Polydispersity would require multiple calculations of $\gamma(r)$, 
as given by Equation \eqn{eq:polyGz} and \eqn{eq:polyG0},
but this would in principle be required to be done only once for 
each $G_0(z)$-spectrum. 
The presence of radial density profiles $\rho(r)$ could be included
already in the Monte Carlo sampling procedure, as described in Appendix 
\ref{appendix:uniform_sampling}. 
For a sphere, this method 
 requires that $r^2\rho(r)$ is described by an analytically
invertible function. If not, an additional sampling 
must be performed, preferrably from a distribution that closely
follows $\rho(r)$, to obtain the desired density distribution. 
Again, this calculation needs to be done only once for each $p(r)$.


\section{Conclusion}

Knowledge of analytical expressions for the SESANS
correlation function $G_0(z)$ for a host of geometries and structures
would be ideal and efficient for calculating model curves and
analyzing experimental SESANS data. As such expressions are hard to
obtain for complex structures due to the difficulty of deriving
general expressions for $\gamma(r)$ to be used in equation
\eqn{eq:GXZ}, one has had to apply the scattering functions from
conventional small-angle scattering to equation \eqn{eq:Gz_SANS}
obtain $G_0(z)$.  
The Monte Carlo algorithm outlined in this paper represents an
alternative method, which is general and 
straightforward to implement from the shape function of the geometry
in question. 
The method 
does not need any of the special functions that
frequently comes with the SANS scattering functions contained in
$\d\Sigma/\d\Omega$.  


%% file: cumuldist.tex
\section{Uniform sampling by the inversion method}
\label{appendix:uniform_sampling}

Standard random number generators provide the user with random real
numbers uniformly distributed between 0 and 1
\cite{NumericalRecipes}. Wishing to sample from a distribution
function $f(x)$ defined on or limited to the interval $a \le x \le b$, 
one sets $f(x)\d x = \d\xi$ and integrates, getting the following
relation between the cumulative distribution $F(x)$ and the random
variable $\xi$:
\begin{equation}
F(x) =  \frac{\int_a^x f(x^\prime)\d x^\prime}
	{\int_a^b f(x^\prime)\d x^\prime} = \xi
\end{equation}
Solving the inverse equation $
%
x = F^{-1}(\xi)$,
%
one can convert the random variable $0 < \xi < 1$ to 
random variables $x$ uniformly distributed in $f(x)$.

Considering a sphere with radius $R$, the three spherical
coordinates $r, \phi, \theta$ have the probability distributions
$3r^2/R^3$, $1/2\pi$, $\frac{1}{2}\sin\theta$, respectively. A
three-dimensional uniformly distributed random variable $\mathbf{\xi}$
is then converted points uniformly distributed throughout the volume
of the sphere by 
\begin{equation}\label{eq:fullsphere}
\begin{array}{lcl}
	r    	   & = & \bracket{\xi_1 R^3}^{1/3} \\
	\phi 	   & = & 2\pi\xi_2 \\
	\cos\theta & = & 1 - 2\xi_3 
\end{array}
\end{equation}
When the geometry in question is, say, a hemisphere, a spherical sector
or a spherical shell defined by the coordinates $\bracket{R_1, \Phi_1,
\Theta_1}$ and $\bracket{R_2, \Phi_2, \Theta_2}$, the distribution
functions can be limited to these regions. 
Equation \eqn{eq:fullsphere} then generalizes to  
\begin{equation}
\begin{array}{lcl}
	r    	   & = & \angular{\xi_1\bracket{R_2^3-R_1^3} +
				R_1^3}^{1/3} \\ 
	\phi 	   & = & \xi_2 \bracket{\Phi_2 - \Phi_1} + \Phi_1\\
	\cos\theta & = & \cos\Theta_1 
	- \xi_3\bracket{\cos\Theta_1 - \cos\Theta_2}. 
\end{array}
\end{equation}
A further generalization is the inclusion of a radial distribution of
scattering length by a function $\rho(r)$. Taking the function
$\rho(r) = r^{-\alpha}$, where $\alpha < 3$, which in particular
applies to spherical star polymers and polymeric micelles
\cite{Halperin}, the radial coordinate $r$ follows the
probability distribution 
$4\pi r^2\rho(r)/V = \bracket{3-\alpha} r^{2-\alpha} / R^{3-\alpha}$, 
yielding in the case of a hollow sphere
\begin{equation}
r = \angular{ \xi_1\bracket{R_2^{3-\alpha} -
R_1^{3-\alpha}}}^{1/\bracket{3-\alpha}}.
\end{equation}
In the case of a cylinder with radius $R$ and length $L$, the
coordinates $r, \phi, z$ follow the respective distributions $2r/R^2,
1/2\pi, 1/L$, giving the relations: 
\begin{equation}
\begin{array}{lcl}
r    & = & \bracket{\xi_1 R^2}^{1/2} \\
\phi & = & 2\pi\xi_2 \\
z & = & \xi_3 L 
\end{array}
\end{equation}
The corresponding generalizations to a cylindrical subvolume are:
\begin{equation}
\begin{array}{lcl}
	r    	   & = & \angular{\xi_1\bracket{R_2^2-R_1^2} +
				R_1^2}^{1/2} \\ 
	\phi 	   & = & \xi_2 \bracket{\Phi_2 - \Phi_1} + \Phi_1\\
	z          & = & \xi_3 \bracket{L_2 - L_1} + L_1.
\end{array}
\end{equation}
For a rectangular box defined by a lower corner
$\bracket{a_1, a_2, a_3}$ and an upper corner $\bracket{b_1, b_2,
b_3}$, the transform relations for the cartesian coordinates
$\bracket{x_1, x_2, x_3}$ takes the simple form
\begin{equation}
x_i = \bracket{b_i - a_i}\xi_i + a_i.
\end{equation}

%% file: mcsesans_figtabs.tex

\clearpage


\begin{figure}
\begin{center}
\includegraphics[width=6.5cm]{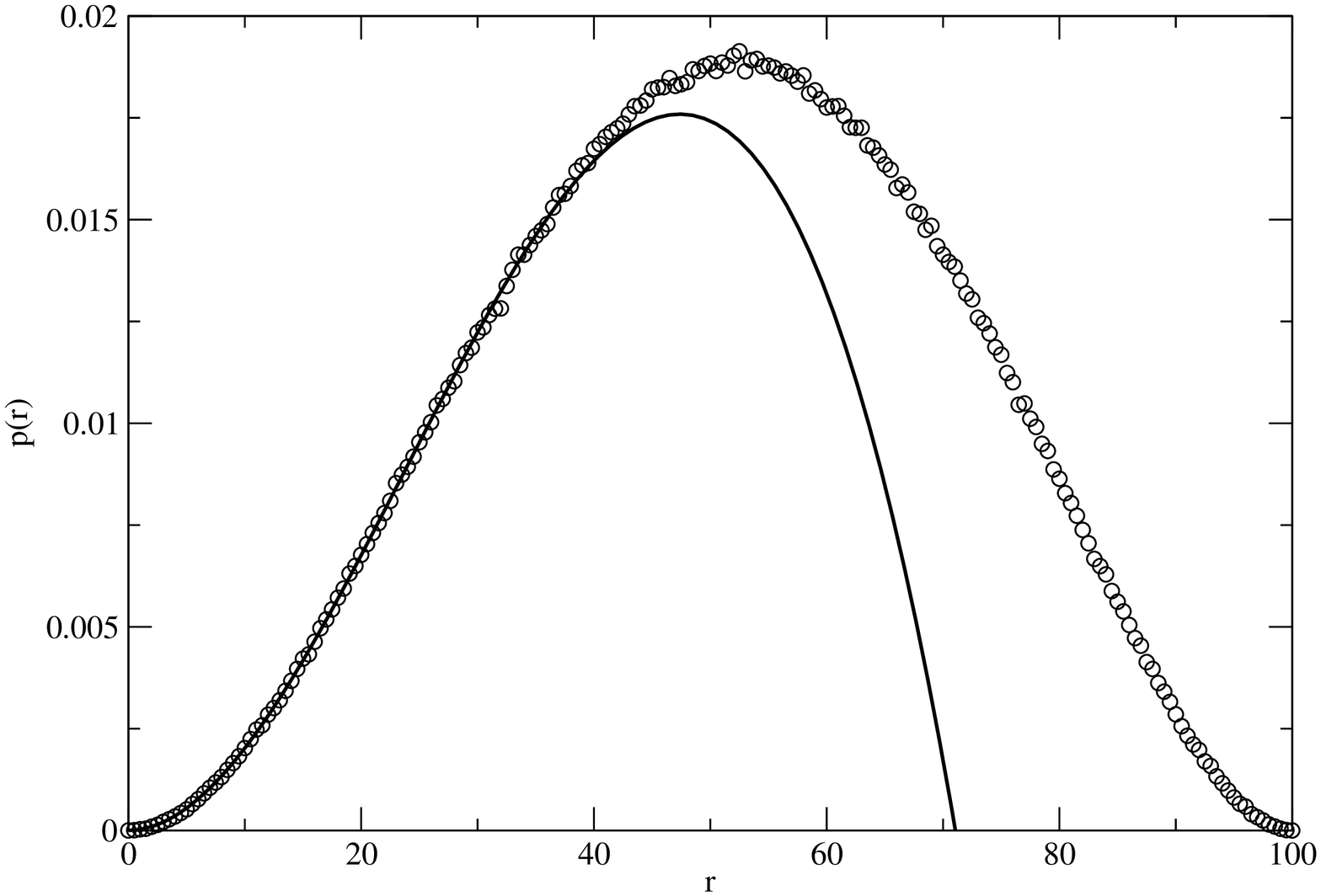}
\includegraphics[width=6.5cm]{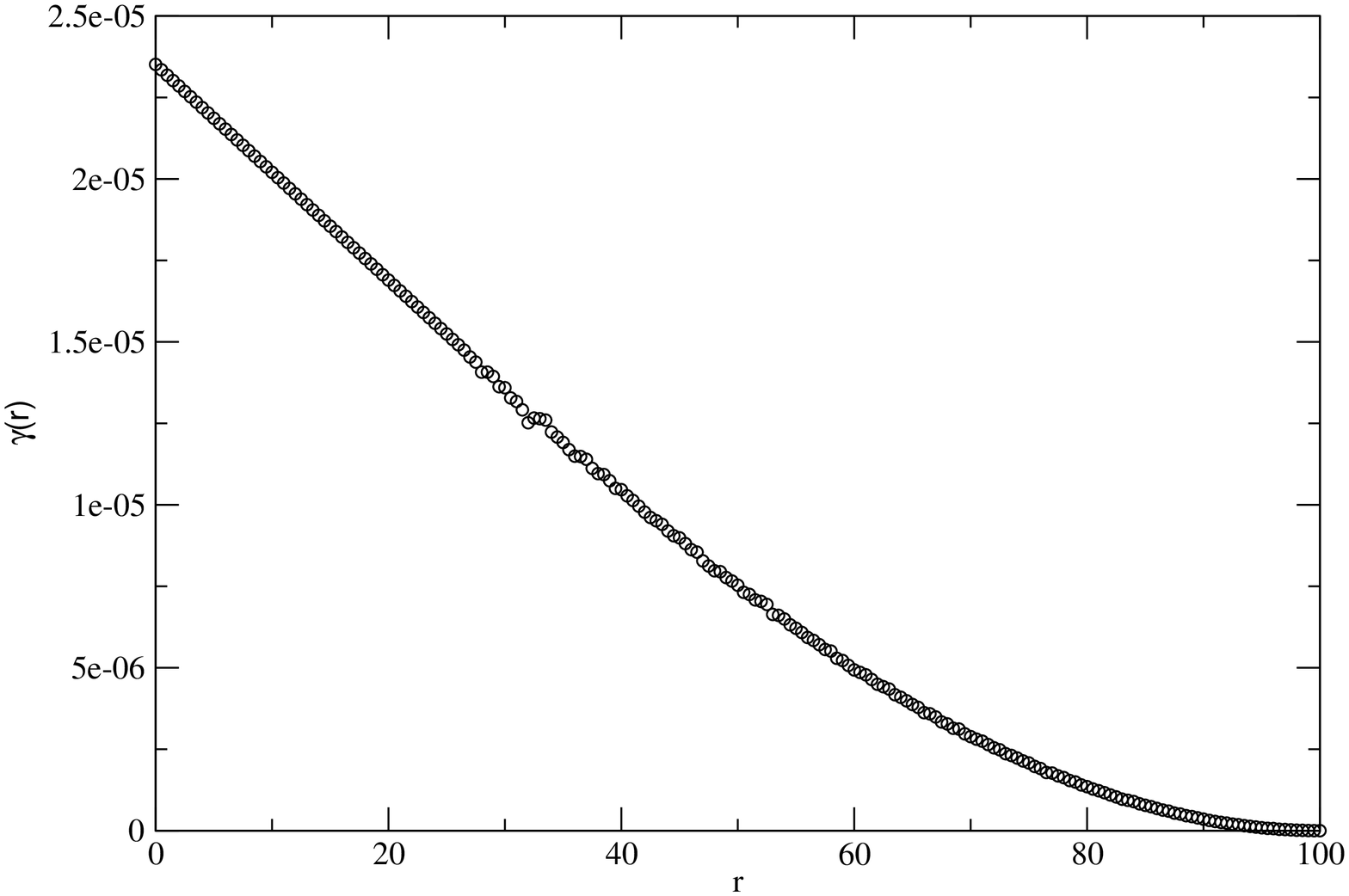}
\end{center}
\caption{Right: Monte Carlo calculated pair distance distribution
function $p(r)$ for a homogenous sphere with radius $R = 50$. The
solid line represents the polynomial $r^2\bracket{a + br +
cr^2}$ that has been fitted to $p(r)$ up to $r = 26.5$. 
Left: density correlation function 
$\gamma(r)$ calculated from the $p(r)$ curve. The low-$r$ part is
calculated from the polynomial fit.}   
\label{fig:Pr_sphere}
\end{figure}

\clearpage

\begin{figure}
\begin{center}
\includegraphics[width=13cm]{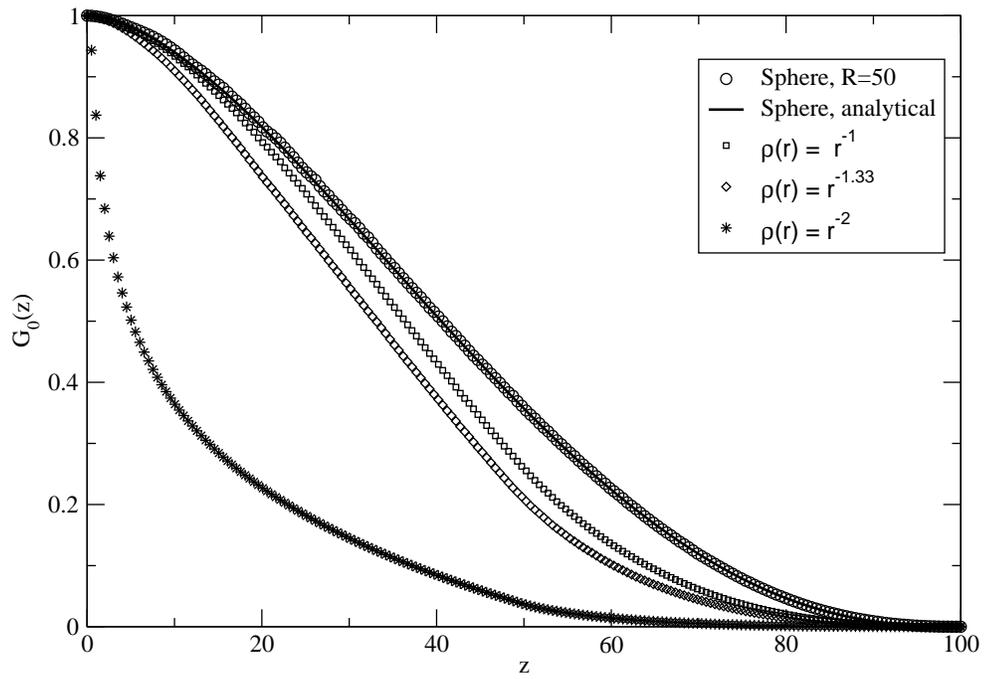}
\end{center}
\caption{Monte Carlo calculated SESANS correlation function $G_0(z)$
for a homogenous sphere with radius $R=50$ (circles).
The solid line represents the analytical expression for $G_0(z)$.
The other curves represent spherical particles with a radial density 
distribution $\rho(r) = r^{-\alpha}$ and a maximum radius $R=50$.}
\label{fig:sphere}
\end{figure}



%

\clearpage

\begin{figure}
\begin{center}
\includegraphics[width=13cm]{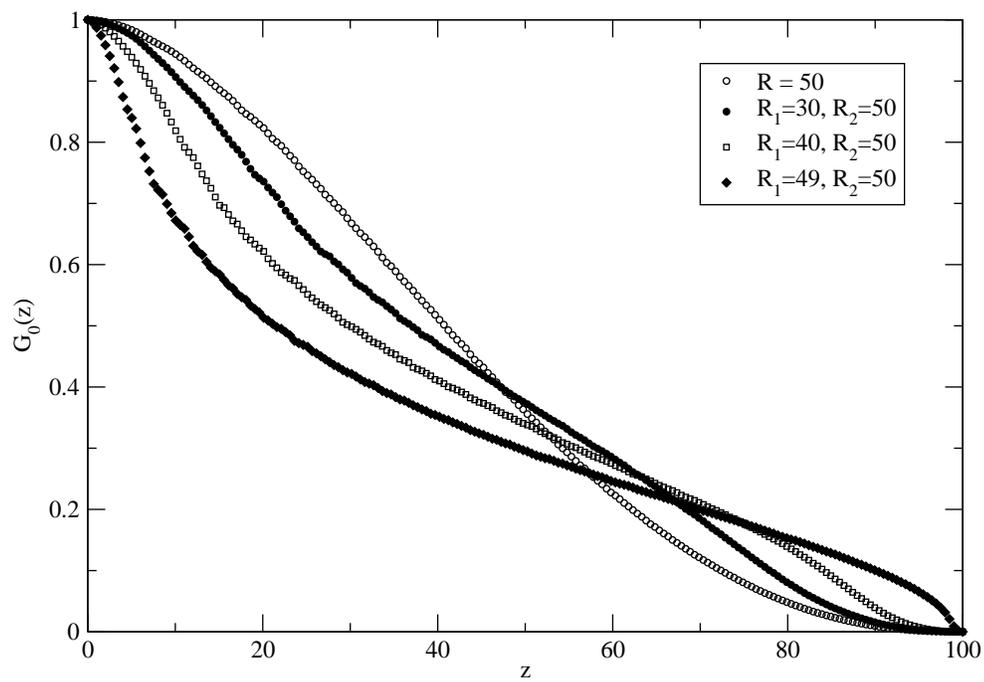}
\end{center}
\caption{Monte Carlo calculated SESANS correlation functions for 
one homogenous (open circles) and various hollow spheres. 
}   
\label{fig:hsphere}
\end{figure}

\clearpage

\begin{figure}
\begin{center}
\includegraphics[width=13cm]{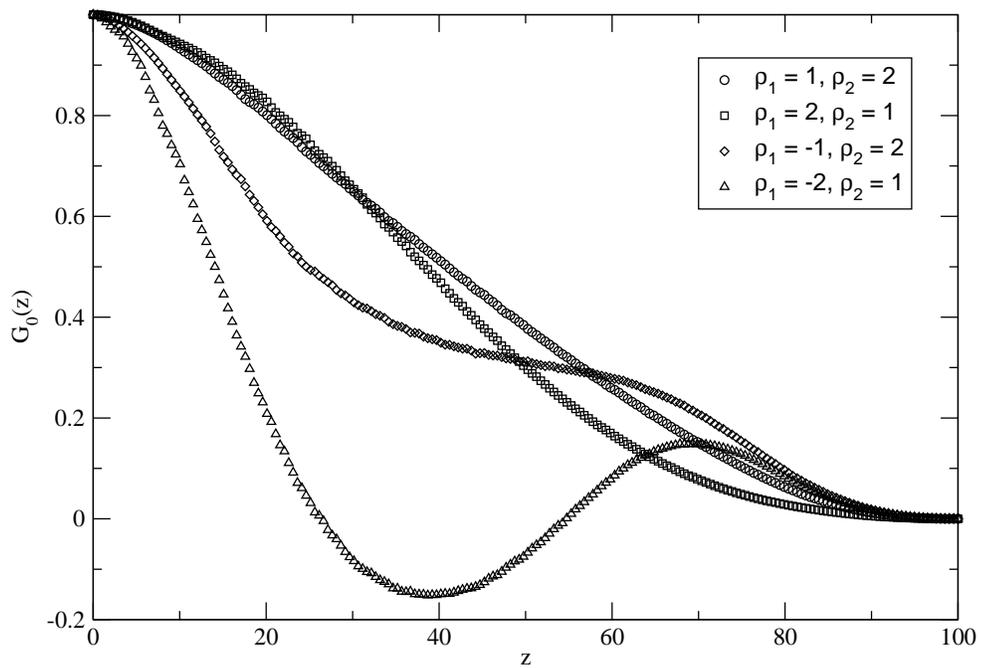}
\end{center}
\caption{Monte Carlo calculated SESANS correlation functions for a
core-shell sphere with inner radius $R_1 = 30$ and outer radius
$R_2 = 50$ with different scattering length densities.}
\label{fig:coreshell_sphere}
\end{figure}

\clearpage

\begin{figure}
\begin{center}
\includegraphics[width=13cm]{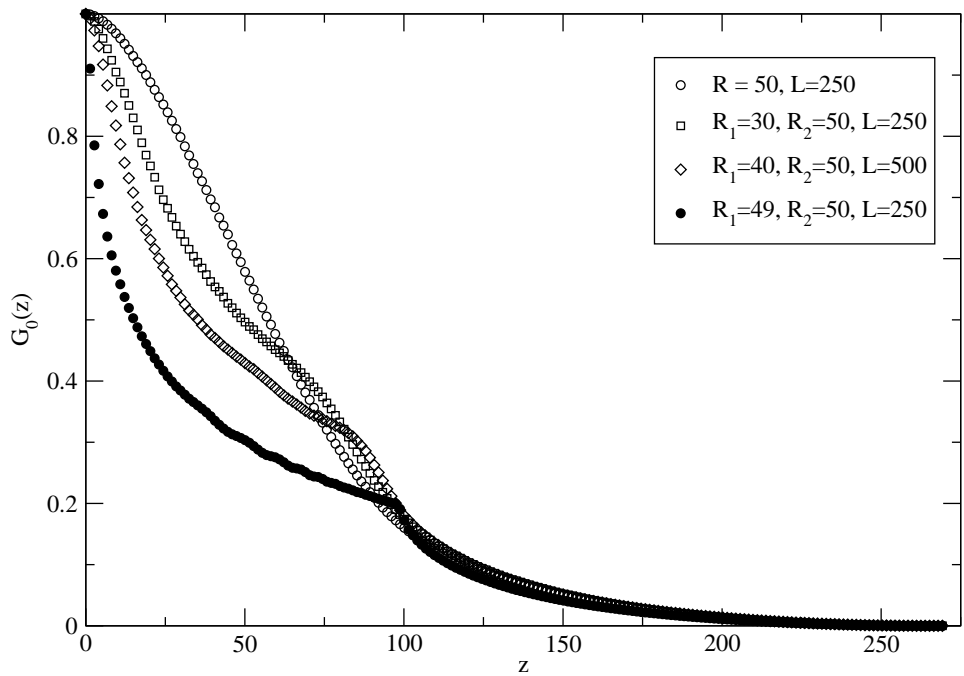}
\end{center}
\caption{Monte Carlo calculated SESANS correlation functions for 
one homogenous and several hollow cylinders with outer radius $R_2 =
50$, different inner radii $R_1$, and length $L=250$.}
\label{fig:cylinder}
\end{figure}

\clearpage

\begin{figure}
\begin{center}
\includegraphics[width=13cm]{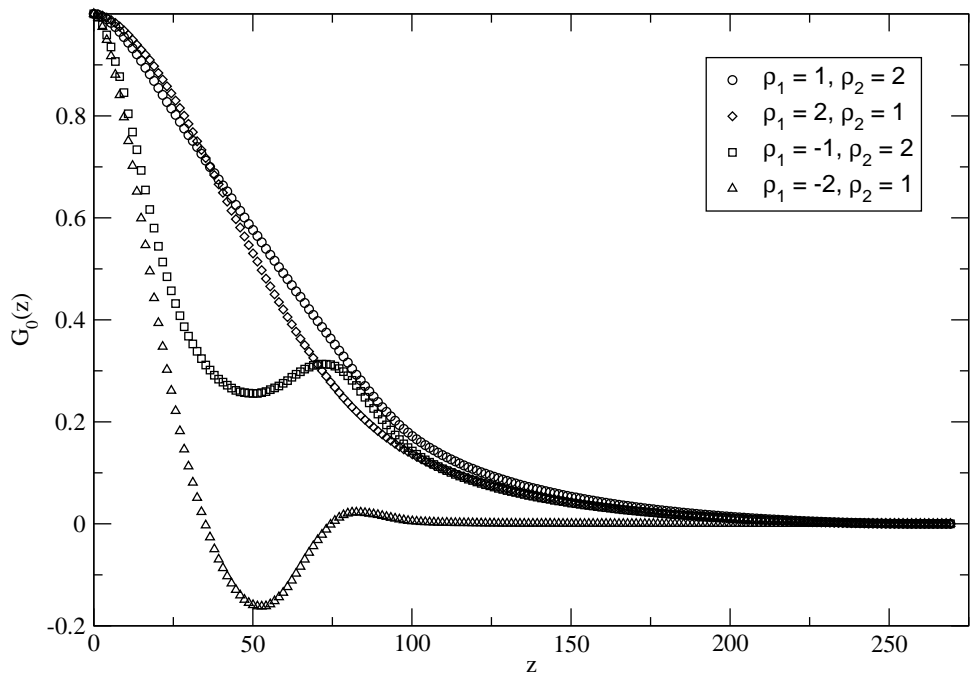}
\end{center}
\caption{Monte Carlo calculated SESANS correlation functions for a
core-shell cylinder with inner radius $R_1 = 30$, outer radius
$R_2 = 50$, and length $L=250$ with different scattering length
densities.} 
\label{fig:coreshell_cylinder}
\end{figure}

\clearpage

\begin{figure}
\begin{center}
\includegraphics[width=13cm]{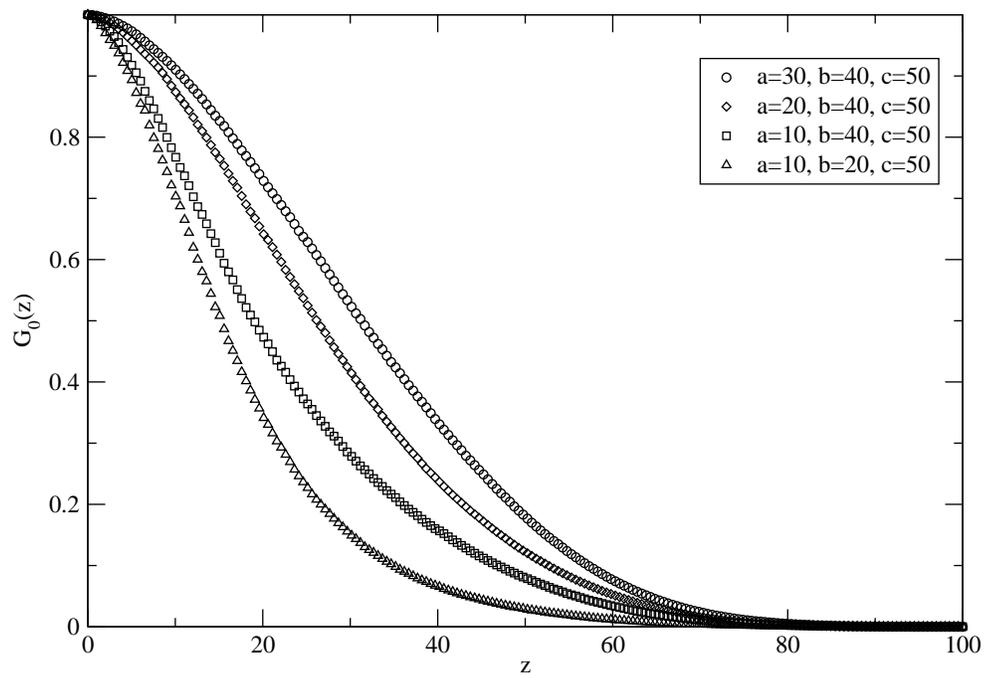}
\end{center}
\caption{Monte Carlo calculated SESANS correlation functions for a
various triaxial ellipsoids.} 
\label{fig:ellipsoids}
\end{figure}